# Intra-Domain Periodic Defects in Monolayer MoS$_2$

Anupam Roy[1*], Rudresh Ghosh[1*], Amritesh Rai[1], Atresh Sanne[1], Kyounghwan Kim[1], Hema C. P. Movva[1], Rik Dey[1], Tanmoy Pramanik[1], Sayema Chowdhury[1], Emanuel Tutuc[1] and Sanjay K. Banerjee[1]

[1]Microelectronics Research Center, The University of Texas at Austin, Austin, Texas 78758, USA
*Address correspondence to: anupam@austin.utexas.edu; rudresh@utexas.edu. A. R. and R. G. contributed equally to this work.

**Abstract**: We present an ultra-high vacuum scanning tunneling microscopy (STM) study of structural defects in molybdenum disulfide thin films grown on silicon substrates by chemical vapor deposition. A distinctive type of grain boundary periodically arranged inside an isolated triangular domain, along with other inter-domain grain boundaries of various types, is observed. These periodic defects, about 50 nm apart and a few nanometers in width, remain hidden in optical or low-resolution microscopy studies. We report a complex growth mechanism that produces 2D nucleation and spiral growth features that can explain the topography in our films.

The many incredible properties of graphene including high carrier mobility (200,000 cm$^2$V$^{-1}$ s$^{-1}$) [1] have made it a very special material both from fundamental science and an engineering point of view. However, the lack of a band-gap in graphene causes high leakage current which makes it unsuitable for many optoelectronic purposes and logic-based devices and circuits. In contrast, transition metal dichalcogenides (TMDs) with the general chemical formula $MX_2$ ($M$ = Mo, W; $X$ = S, Se, Te) provide a large family of two-dimensional (2D) crystals that vary greatly in physical and chemical properties [2], ranging from metallic to semiconducting to insulators. Of all the TMDs, molybdenum sulfide (MoS$_2$), with its indirect-to-direct band gap transition as a function of layer thickness, has been of particular interest for digital and optoelectronic applications. MoS$_2$ has already been used to fabricate functional electronic circuit elements [3-6], as well as used for optoelectronics [7-9], valleytronics, spintronics [10, 11] and coupled electro-mechanics [12].

Most of the MoS$_2$ material characterization and device demonstrations so far have been on exfoliated samples which suffer from low yield, and cannot be scaled up for practical applications. In order to address these problems, significant work has been done to introduce different growth techniques. Processes including liquid exfoliation [13] and direct sulfurization of molybdenum thin films [14] have been achieved to synthesize large MoS$_2$ monolayers. However, the overall simplicity and the high quality of films obtained using the sulfurization of MoO$_3$ has made it one of the most widely used methods of synthesizing large area monolayer MoS$_2$ [15-17].

Just like different synthesis techniques, various analytical techniques have been introduced. In addition to the commonly used techniques like scanning electron microscopy (SEM) and atomic force microscopy (AFM), techniques like Raman and photoluminescence (PL) spectroscopy have become common to ascertain the number of layers of these 2D materials. However, because of the resolution limit, these techniques only reveal a partial picture. SEM and AFM together show us the topographical and structural information. The spectroscopic techniques ascertain the energy levels to a certain degree. Recent techniques like microwave impedance microscopy (MIM) [18] have been used to map the dielectric constant of these films. However, most of these techniques lack the ability to image these films in the range of a few nanometers, the scale necessary for properly characterizing defects.

Scanning tunneling microscopy (STM) is one of the few techniques that can image the surface of a material to the scale of a few nanometers. Microscopic studies of surface defects in monolayer $MoS_2$ have been investigated mostly by scanning/transmission electron microscopy (S/TEM) [16, 17, 19-23]. There have been a few studies using STM for defect metrology [24-27], but the technique has not received much attention, in part due to the difficulty of probing exfoliated samples. Initial exploration on exfoliated $MoS_2$ surface showed naturally occurring surface defects which give rise to a variation in the local stoichiometry of $MoS_2$ and can be correlated with the resultant Schottky barrier height in metal-$MoS_2$ contacts, as well as the *n*- and *p*-type behavior of the material [26, 27]. However, they concluded that a large variation exists between different samples of exfoliated $MoS_2$ and that is necessary to find a synthesis route for obtaining high quality $MoS_2$. So far, STM characterization of the defects in CVD grown $MoS_2$ was mostly confined to the study of point defects and grain boundaries in rotationally commensurate epitaxial graphene/$MoS_2$ system [28, 29] and spiral growths [30].

As seen in multiple studies, the CVD grown films have shown uniformity at large scale as well as good material behavior in terms of crystallinity. However, electronic devices based on CVD $MoS_2$ have consistently performed poorly compared to the devices based on exfoliated flakes. This indicates that CVD $MoS_2$ may contain more intrinsic defects, *e.g.*, point defects. To understand the origin of the below-par performance, we characterized the film by STM at room temperature (RT) and, surprisingly, found that in addition to the point defects, as in exfoliated flakes, other major surface defects exist on CVD-grown $MoS_2$ films.

$MoS_2$ films for this study were grown on silicon substrates using sulfurization of molybdenum oxide ($MoO_3$) using a standard vapor transport growth process. The $MoO_3$ and sulfur powders were placed in alumina crucibles and loaded in a single zone horizontal tube furnace along with the Si(100) growth substrates. The growth was done at 850 ºC for 5 min following a cooldown to RT. More details of the growth can be found in Ref. 5.

Post-growth investigations were carried out using a Renishaw inVia Raman microscope where a 532 nm diode laser was used for the Raman (3600 *l/mm*) and PL (1200 *l/mm*) measurements. The STM measurements were carried out using a tungsten tip at RT under ultra-high vacuum (base pressure ~ $2\times10^{-10}$ mbar). Details of the system have been described elsewhere [31]. Chemical stoichiometry of the film was further confirmed by XPS using a monochromatic Al-K$\alpha$ source ($hv$ =1486.7 eV) operating at 15 kV. The plan-view TEM samples were prepared by transferring the as-grown film onto holey carbon TEM grids using a poly (methyl methacrylate)-based wet-transfer technique and the TEM images were taken using a FEI-Tecnai TF20 microscope.

Figure 1 shows the SEM images of the deposited film at various distances from the $MoO_3$ precursor. Far from the $MoO_3$ precursor, the local concentration of the vapor phase material is expected to be low. As a result, there are fewer nucleation sites and the individual domains are reduced in size [Fig. 1(a)]. As it gets closer to the $MoO_3$ source, the flakes become larger in size with more number of domains appearing inside the same area [Fig. 1(b)]. As the size of individual triangle reaches 30~50μm, neighboring grains start to merge [Fig. 1(c)] and eventually form a pseudo-continuous film [Fig. 1(d)]. This pseudo-continuous film usually spreads over a large area (~ a few $mm^2$). Chemical stoichiometry of the film is investigated through X-ray photoelectron spectroscopy (XPS) [See supplementary material Fig. S1 (a) & (b)]. Furthermore, AFM measurement confirms monolayer thickness [See supplementary material Fig. S2 (d) & (e)].

Raman spectra from the as-grown sample shows the two dominant peaks ($E^1_{2g}$ and $A_{1g}$) separated by 19.5 $cm^{-1}$ which is characteristic of CVD-grown monolayer $MoS_2$ [13-17]. The material quality of the film can be correlated to the full width half maximum (FWHM) of these peaks. For the as-grown $MoS_2$ samples we measure a FWHM of 6.69 $cm^{-1}$ for the $A_{1g}$ peak and 5.3 $cm^{-1}$ for the $E^1_{2g}$ peak, which is comparable to exfoliated monolayer $MoS_2$ flakes. A map of the Raman peak intensity also confirms the uniformity of the grown film [See supplementary material Fig. S2 (a)]. PL spectrum from the same region shows a dominant peak at around 678 nm (1.83 eV), which is also similar to what is expected for monolayer MoS$_2$. The Raman and PL spectra before and after the transfer of MoS$_2$ from the growth substrate onto a SiO$_2$/Si substrate are shown in Fig. 2. The difference in spectra between the as-grown and transferred MoS$_2$ samples has been explained in terms of the inherent strain that occurs during the growth process [21, 32-34]. The hexagonal surface structure examined by plan-view TEM [See supplementary material Fig. S1 (c) & (d)] mostly shows no defects, indicative of the film attaining a relaxed configuration after being transferred onto another substrate. However, traces of some of the defects from the as-grown film can be found [See supplementary material Fig. S1 (e) & (f)]. Another noticeable change is observed in the $A_{1g}/E^1_{2g}$ peak intensity ratio that can arise from substrate-induced interference effect due to the difference in size of MoS$_2$ domains and/or the roughness of the grown film [34]. The transfer process can reduce the strain in the grown film, thereby relaxing the film. This can also bring changes in the overall roughness of the film as the transfer process can lead to the healing of several defect lines.

In Fig. 3, we present RT STM study of MoS$_2$(0001) grown on Si(001) surface by CVD. Figure 3 (a) shows an isolated triangular domain of MoS$_2$ film similar to the regions shown in Fig. 1 (a) and (b). Interestingly, STM studies show a very different picture than is expected from our previous imaging techniques of a single domain. Unlike the continuous domain observed by optical, electron, or atomic force microscopy, the individual domain in reality appears to be built of several concentric triangles. These periodic surface defects at the nanoscale are unlike GBs as they appear inside individual domains. They are visible over all scanned regions and do not seem to be the result of any localized effects that might be due to substrate impurities. We refer to these defects as intra-domain periodic defects (IDPDs). These IDPDs are about 50 nm apart from each other with the average width of each IDPD being about 10 nm under the STM bias condition used here, and therefore, may not be observed by SEM or AFM. The height profile drawn across many IDPDs on the domain (black line), shown in Fig. 3 (e), confirms that the whole domain is of the same height, as expected of an individual monolayer domain. Each IDPD corresponds to a drop in the height of ~ 0.3-0.4 nm as observed in the present scan condition, which is close to half a monolayer thickness of MoS$_2$. The presence of such defects can change the resistance and can lead to larger leakage currents through the grain boundaries. Examination of multiple samples by STM shows the presence of such concentric triangles as the building blocks of individual larger domains that, in turn, are the building blocks of pseudo-continuous films that stretch over many mm$^2$.

In Fig. 3 (b) and (c), we see multiple individual domains oriented in random directions merge together to form a continuous film. All the smaller domains are triangular in shape, reflecting the hexagonal structure. However, when they merge together, different type of defects in the form of GBs arise. The GBs, marked by green and black arrows are different from the IDPD, shown by red arrow in Fig. 3 (b), and are localized in the region where different domains merge together. Figure 3 (b) data shows that this merging can occur in different ways. One of the merging types is the mirror twin [center domains of Fig. 3 (b) shown by black arrow, and also in supplementary material Fig. S3 (a)] that is characterized by two individual triangular domains meeting at 180 degrees to one another, and has also been observed in TEM studies by several other groups [16, 35]. The second type of defect shown is the 'tilt twin' where the merging happens at a random angle [GBs shown in green in Fig. 3 (b), and also in supplementary material Fig. S3 (a) & (b)].

When a sample starts to crystallize in CVD growth, depending on the growth conditions, many seeds can be formed and each seed grows until they meet at the boundaries. The size of the triangular domain can vary depending on the number of neighboring nucleation seeds. As can be seen in Fig. 3 (c), all of these triangular domains show the presence of IDPDs, and the separation between two IDPDs depends on the proximity of other domains [also in supplementary material Fig. S3 (a) and (b)]. Depending on the edge spacing between two separate domains, the spacing between GBs will be different. Although for the isolated domain the concentric triangles appear to be strictly equilateral in shape, the sizes and shapes of domains in regions where these individual domains merge are dictated by the sizes and shapes of merging domains.

The growth of a crystal, in general, is explained based on three principal mechanisms – layer-by-layer (LBL), screw-dislocation-driven (SDD) spiral, and dendritic growth mechanisms [36-39]. SDD growth mechanism is preferred at lower supersaturation where threading dislocations, such as slipped planes at the basal domain with a screw component present on the surface providing a continuous step source, can lead to a spiral-like growth. Spiral-like growth has been predicted by Burton, Cabrera, and Frank (BCF theory) in their pioneering theoretical work [38], and has been observed for the different transition metal chalcogenide systems [30, 40-42]. Presence of such spiral-like growth on a triangular domain is shown in Fig. 3 (a) & (d), and also in supplementary material Fig. S3 (c). Spirals of both types - rotating clockwise and counter-clockwise - are observed [Fig. 3 (d) and supplementary material Fig. S3 (c)]. Fig. 3 (f) shows the line profile along the green line drawn across the spiral in Fig. 3 (a) with a monolayer step height (~ 0.8 nm). Similarly, line profile [Fig. 3 (g)] taken along the blue line in Fig. 3 (d) also shows spiral stacking with about a monolayer step height.

Figure 3 (d) shows a domain that contains both the 2D monolayer nanosheet with concentric IDPDs and spirals at the center where it splits into two triangular domains. As shown in the schematic [inset, Fig. 3 (d) and Fig. S4], many of the triangular domains show the film winding up mostly to the bilayer height at the center of the domain due to screw dislocations, whereas, away from the center the domain spreads as a monolayer 2D nanosheet with concentric IDPDs. The supersaturation of the system during growth drives the crystal to adopt a particular growth mechanism. Dendritic growth dominates at higher supersaturation, whereas, lower supersaturation growth condition promotes SDD growth [36-39]. The intermediate condition is suitable for LBL growth, where critical supersaturation condition on both sides can lead to SDD growth for lower and dendritic growth for higher supersaturations. Controlling the growth condition on various substrates can lead to different growth mechanisms and many interesting thin film morphologies [43, 44]. A combination of low supersaturation and high growth temperature can promote a complex growth mechanism, as observed in our case, where both large 2D nucleation growth and dislocation driven spirals occur – a signature that both LBL and SDD growth modes may be at play.

To understand the possible origin of GBs and IDPDs formation, we suggest that the CVD growth of $MoS_2$ thin films at a temperature as high as 850 °C and on different substrates could produce a different film-substrate coupling, depending on the thermal expansion coefficients of the film and the substrate. The difference in thermal expansion coefficient plays a crucial role in the material quality of the grown layer, and can introduce strain in the grown film when cooled down from high temperature to RT, acting as a driving force for the formation of defects such as GBs, IDPDs etc. This is not unusual, as there are several reports that show non-uniform mapping of strain in CVD-grown single-crystalline monolayer $MoS_2$ [21, 33, 45-47]. The difference in the thermal expansion coefficient of the thin native $SiO_2$ layer and Si can lead to a tensile stress gradient in the $SiO_2$ layer [48]. Thus, at high temperature, the surface can relax the strain by forming various defects which can act as reaction centers. Furthermore, a nearly six times difference in the thermal expansion coefficients of $SiO_2$ and $MoS_2$ layers [21, 49] can create a significant lattice mismatch during a fast-cooling process from the growth temperature (~ 850 °C). This can add a significant contribution to the formation of defects and different types of GBs and IDPDs, as evident from the STM studies [Fig. 3], which, in turn, can potentially impact the electrical properties of the film [16, 17, 50]. Moreover, the missing atoms along an IDPD produce a large number of broken bonds, and these sites may be more prone to the adsorption of impurities when exposed to atmosphere as this is favorable from a thermodynamic point of view to lower the total free energy of the surface [51]. The growth condition, substrate, and post-growth treatment play a significant role on the film morphology as well as on the film-substrate interaction, and are therefore worthy of further investigation.

In conclusion, we studied $MoS_2$ films grown on Si substrate using CVD. Low-resolution microscopy studies show large area, uniform monolayer $MoS_2$ triangular domains. STM studies conducted on the same film shows the presence of defect lines that are a few nanometers wide, half a monolayer in depth, and arranged periodically in a concentric manner on an isolated triangular domain.

The STM studies show 2D nucleation in conjunction with spirals, suggesting that there exists a complex growth mechanism.

**Acknowledgements**: This work was supported in part by NRI SWAN and NSF NNCI. We appreciate technical support from Omicron.

**Supplementary Material**: See supplementary material for the additional characterization data and the growth mechanism: S1: XPS and TEM Characterization, S2: Raman Mapping, PL and AFM Characterization, S3: Defects in As-grown $MoS_2$ Probed Through STM, S4: Growth Mechanism.

**Figure 1**

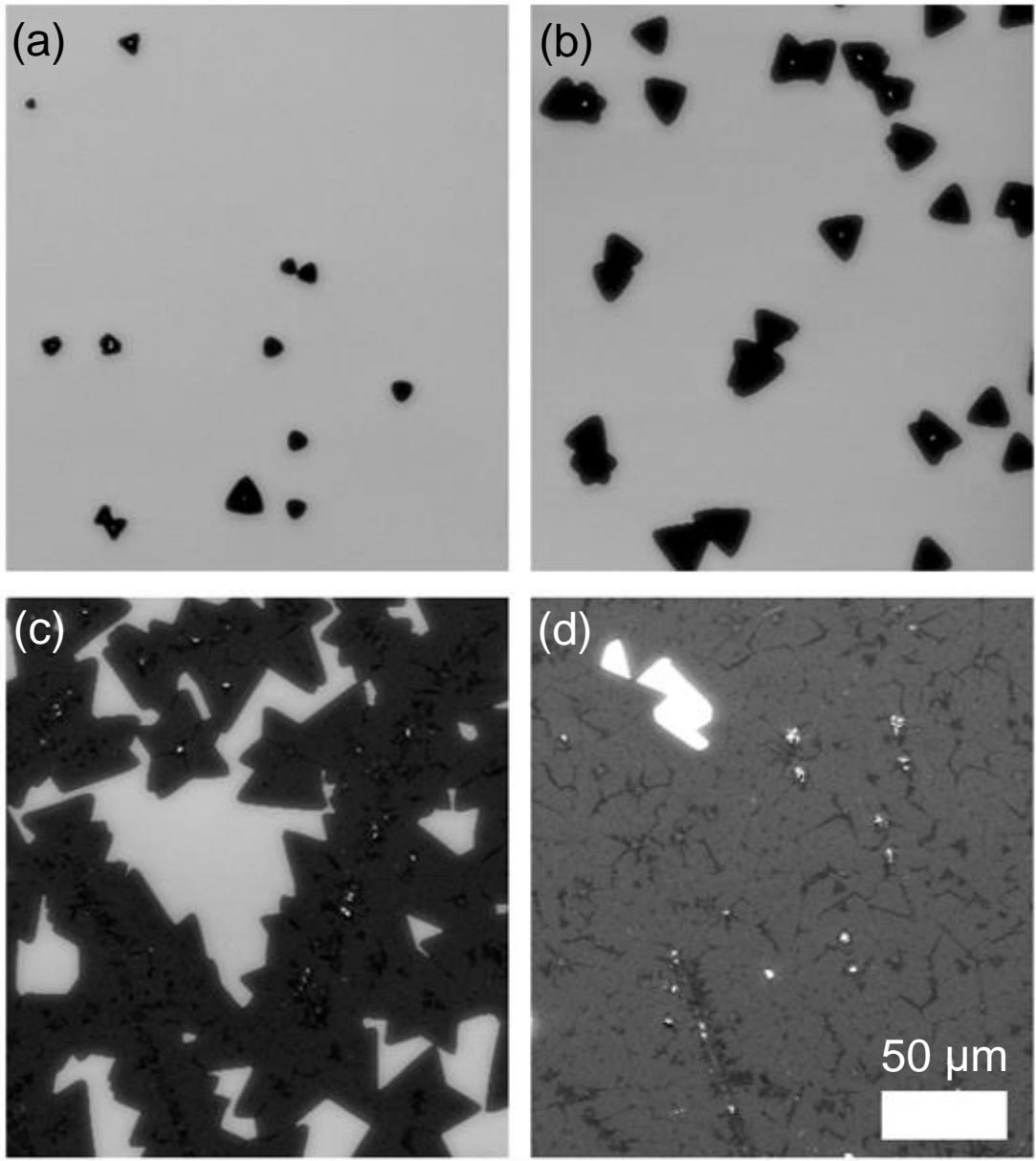

**Figure 1**. SEM images show the change in domain size and nucleation density as a function of the distance from the MoO$_3$ precursor [(a) 31 mm, (b) 30 mm, (c) 29 mm and (d) 28 mm from MoO$_3$ precursor]. The edge-lengths of isolated domains vary from less than 10 μm to larger than 50 μm, beyond which the individual domains merge to form pseudo-continuous films. All the images are at the same magnification.

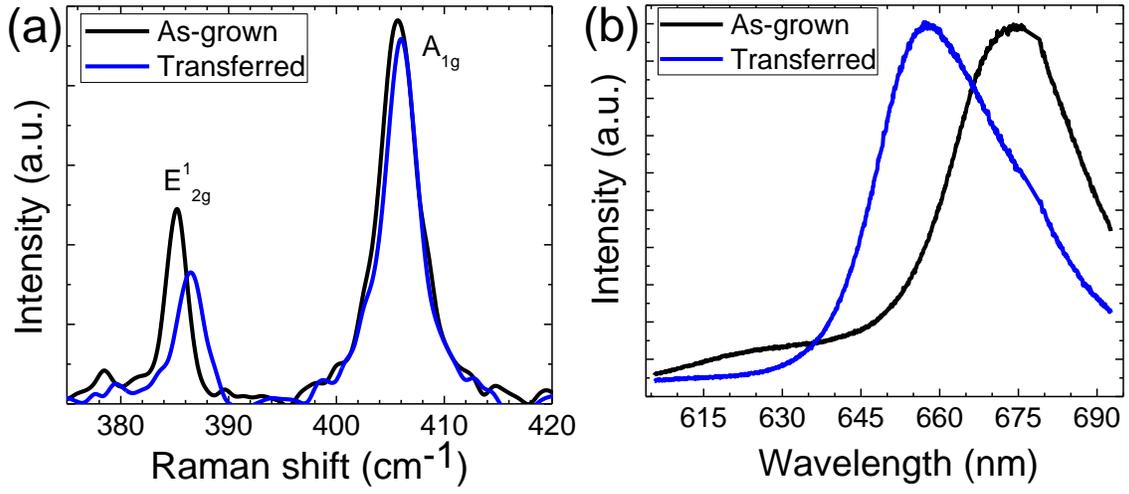

**Figure 2**. (a) Raman spectra from the as-grown and transferred MoS$_2$ films showing a shift in characteristic peak position. (b) Strain-induced energy shift in the photoluminescence spectra from the same sample before and after transfer onto SiO$_2$/Si substrate.

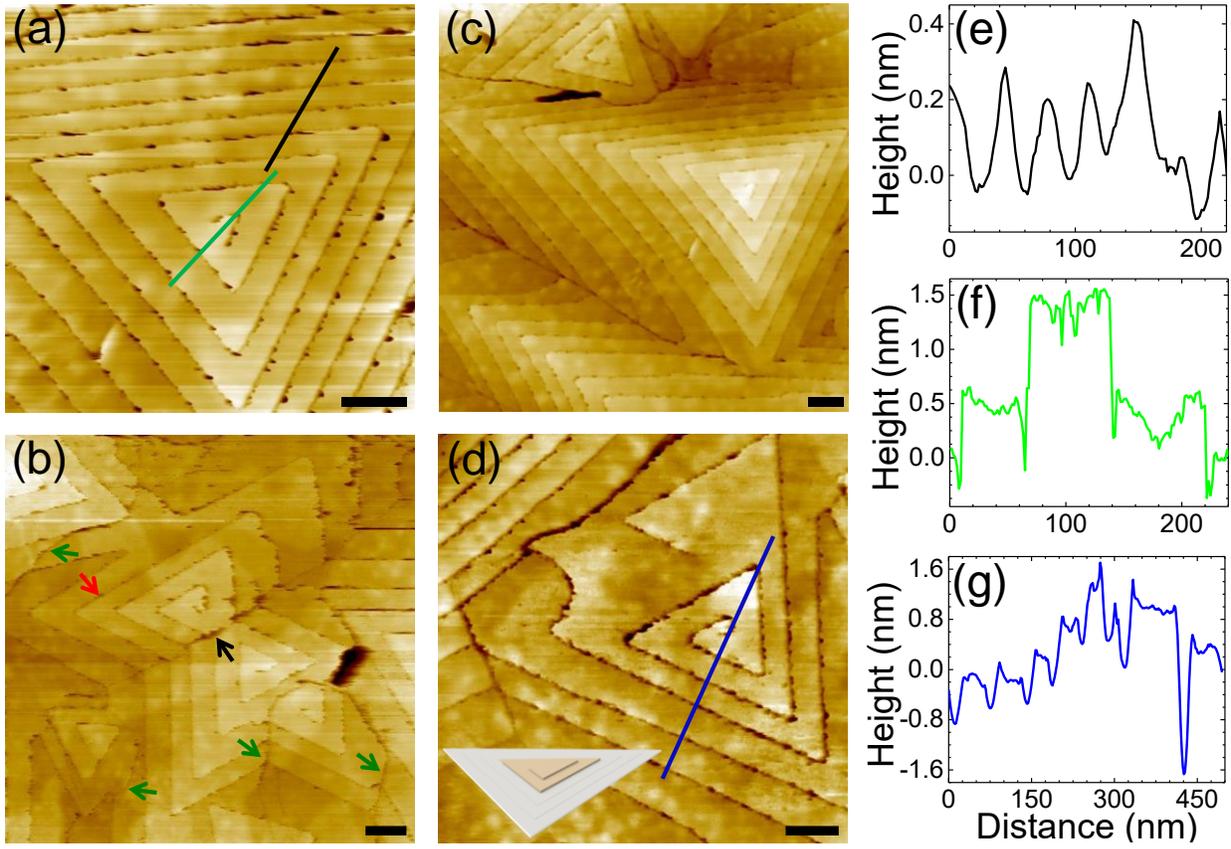

**Figure 3**. STM study at RT in the constant current mode on MoS$_2$ thin films grown on Si(001) surfaces by CVD. For all the images, the bias voltage and the tunneling current are 2.4 V and 0.7 nA, respectively. (a) An individual triangular domain shows concentric triangular grains separated by IDPDs. (b) Triangular domains merge together in random orientation showing various defects. (c) Growth of triangular domains is restricted by neighboring domains and thus, shows different IDPD spacing. (d) STM image showing spiral growth in addition with IDPDs arranged as concentric triangles. Inset shows a schematic of a triangular domain consisting both IDPDs and spirals. Height profiles on (a) showing all GBs on the same plane of a monolayer domain (e, black line) with the spiral-like bilayer height at the center of the triangle (f, green line). Height profile on (d) showing a spiral growth originating at the intersection of two triangular domains (g, blue line). All scale bars are 100 nm.

# Supplementary Material

**S1: XPS and TEM Characterization**

Fig. S1 (a) and (b) show high-resolution X-ray photoelectron spectroscopy (XPS) scans of the Mo-3$d$ and S-2$p$ peaks from the as-grown sample, respectively. The XPS peak positions are consistent with values reported in literature [1]. The as-grown sample when transferred to a transmission electron microscopy (TEM) grid shows a hexagonal lattice as expected for $MoS_2$ [Fig. S1 (c) and (d)]. The inset of Fig. S1 (c) shows the selected area electron diffraction pattern of the film taken along the [001] zone-axis, similar to prior studies [2-4]. The periodic defect lines are however, not seen in the plan-view TEM images, possibly due to the healing of defects after transferring the as-grown strained film. However, presence of some of the isolated defects are observed in Fig. S1 (e) and (f).

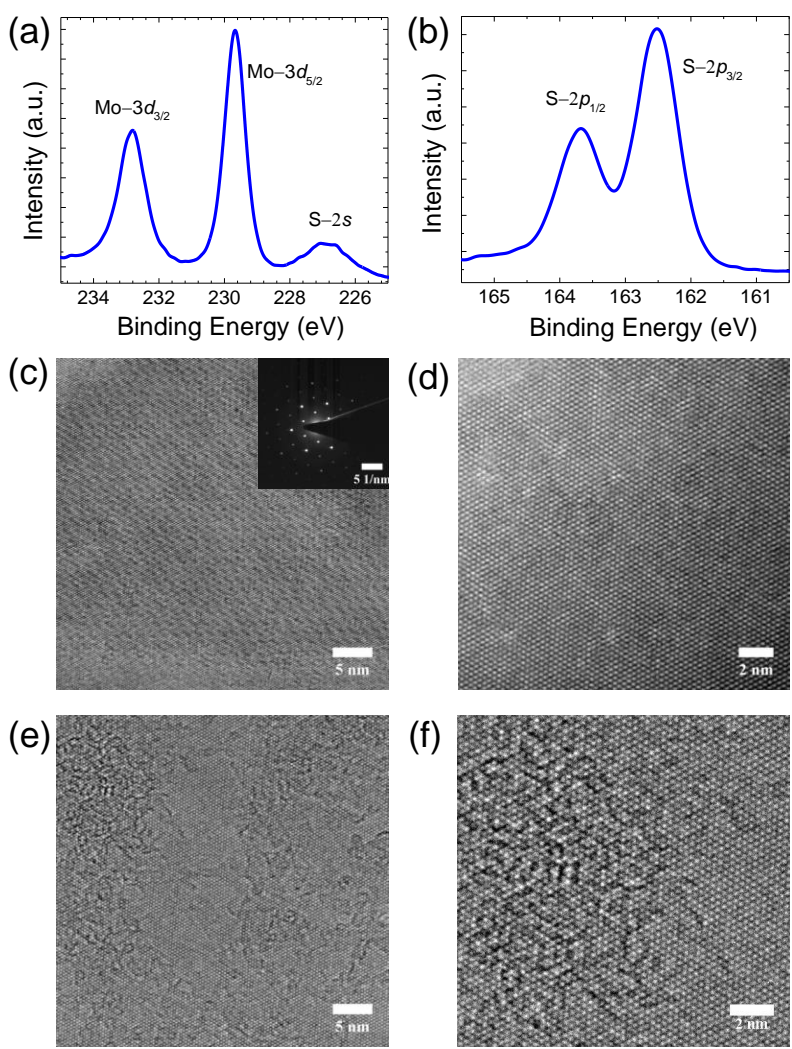

**Figure S1**: High-resolution XPS spectra from the as-grown $MoS_2$ monolayer film. Peak positions corresponding to (a) Mo-3$d$ and (b) S-2$p$ spectra are consistent with 2$H$-$MoS_2$ thin films. (c-d) Plan-view TEM images of $MoS_2$ films transferred onto a TEM grid show the hexagonal surface arrangement of atoms. (e-f) Plan-view TEM images showing presence of some of the isolated defects.

## S2: Raman Mapping, PL and AFM Characterization

Fig. S2 shows the results from Raman and Photoluminescence (PL) experiments conducted on the as-grown films. Atomic force microscopy (AFM) scans show uniform monolayer height of the as-grown sample. The results are consistent with previous studies [4].

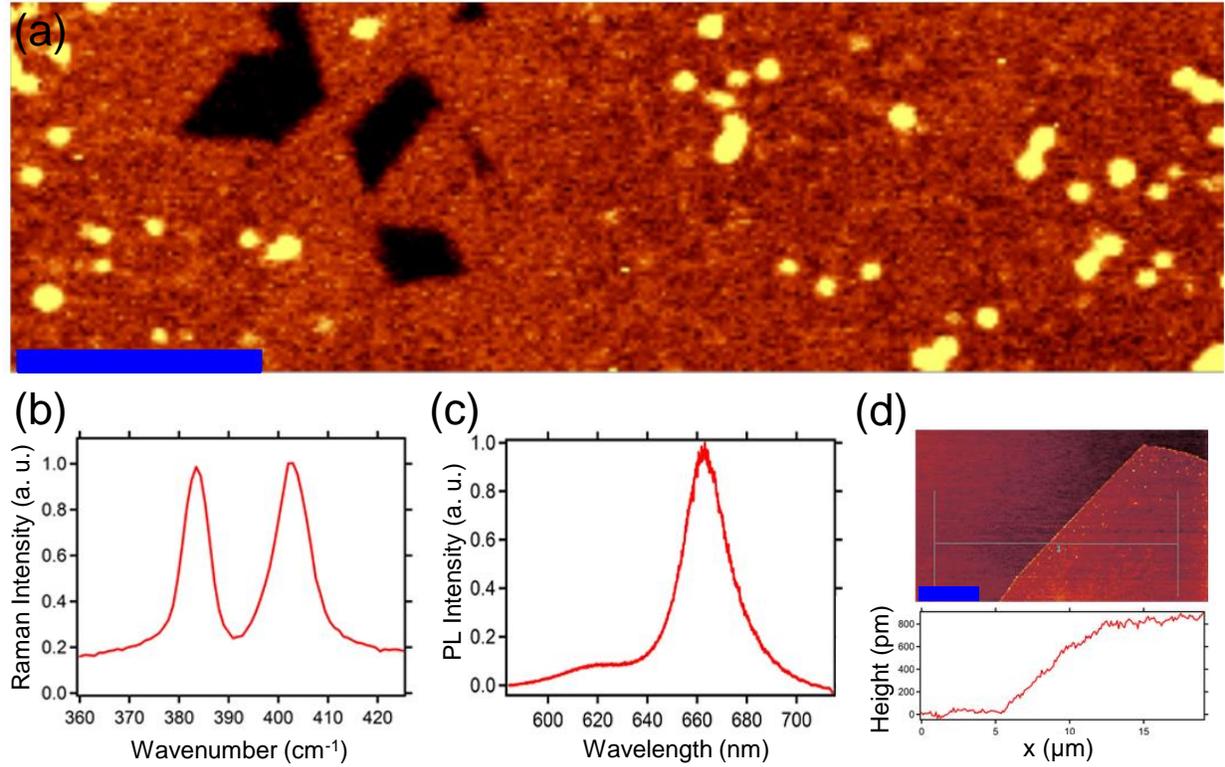

**Figure S2**: (a) The Raman map corresponding to the $E^1_{2g}$ and $A_{1g}$ peaks (sum of total counts between 375 cm$^{-1}$ to 390 cm$^{-1}$ and between 392 cm$^{-1}$ and 415 cm$^{-1}$) for a pseudo-continuous monolayer MoS$_2$ film. The dark regions correspond to uncovered areas. The signal from these regions correspond to those seen for bare Si, while the bright spots correspond to ad-layers of MoS$_2$ which show higher Raman counts than the monolayer. The scale bar is 30 μm. (b, c) Raman and PL spectra from representative points of panel (a). (d) AFM image and the line profile of the marked region of the film confirming monolayer thickness. The scale bar is 5 μm.

## S3: Defects in As-grown MoS$_2$ Probed Through STM

Fig. S3 shows STM images of the as-grown monolayer MoS$_2$ film taken at room temperature. Fig. S3 (a) and (b) shows triangular domains of various sizes merging together. Various types of grain boundaries are observed. Fig. S3 (c) shows a triangular domain with in-plane periodic defect lines arranged as concentric triangles as well as spirals. A height profile of one monolayer across the line drawn in (c) is shown in Fig. S3 (d).

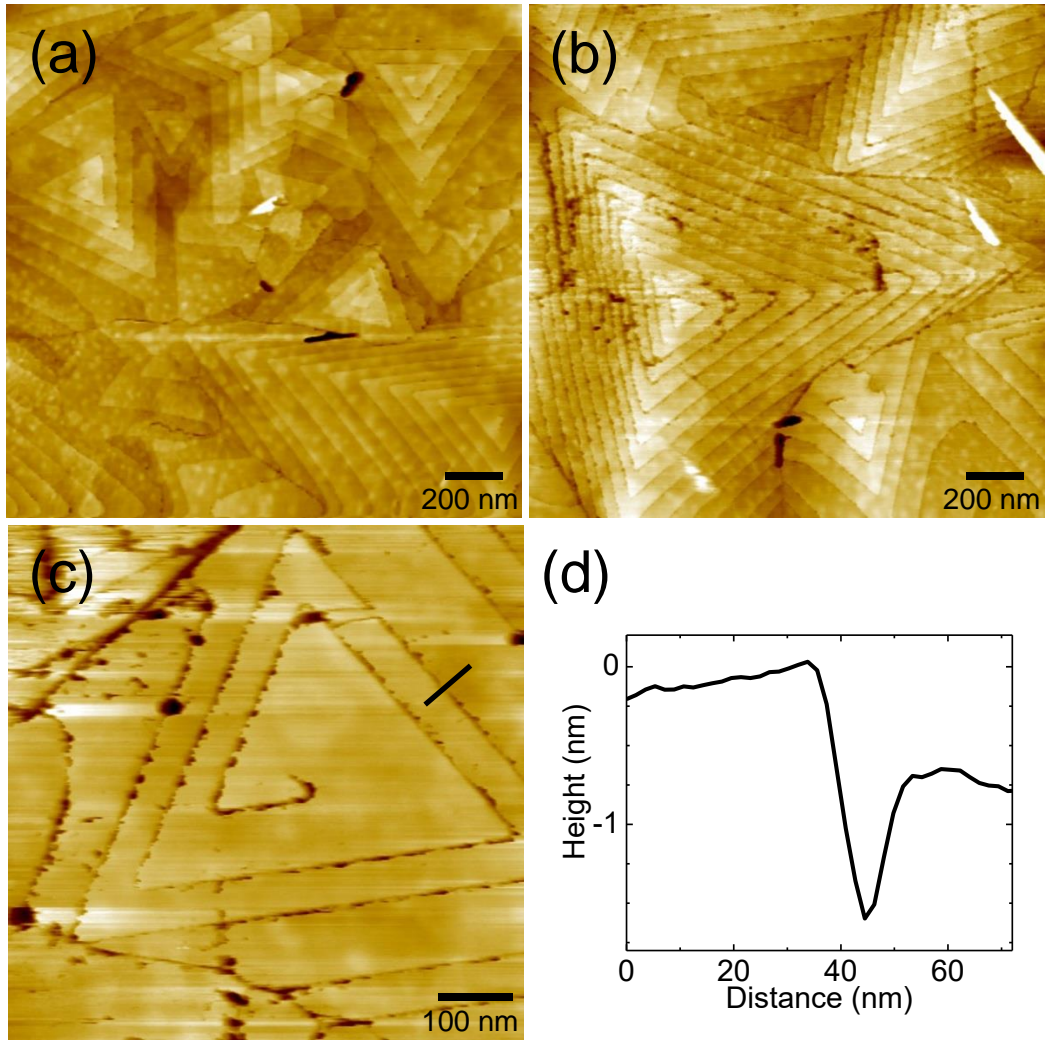

**Figure S3**: Room temperature STM images taken in the constant current mode on CVD-grown monolayer MoS$_2$. The bias voltage and tunneling current are 2.4 V and 0.7 nA, respectively, for all images. (a) Triangular domains merge together in a random orientation showing various defects. (b) STM image showing growth of triangular domains in random directions are restricted by neighboring domains and as a result shows different spacing between defect lines. (c) A triangular domain shows spiral-like growth. (d) A height profile drawn across the step in (c) showing one monolayer height.

## S4: Growth Mechanism

Fig. S4 presents the growth mechanism schematically. Both 2D nucleation and spirals are present. Fig. S4 (a) shows the schematic of the growth where the film piles up during growth as a spiral due to dislocation with a screw component. Fig. S4 (b) shows both the spirals at the center and 2D nucleation away from the center with periodic defect lines in a concentric triangular form. As presented in the schematic in Fig. S4 (c) and the STM image in Fig. S4 (d), these periodic defect lines are on the same plane, whereas the spirals show an increase of monolayer height. The height profile shown in Fig. S4 (e) drawn across the line in Fig. S4 (d) agrees very well with the schematic.

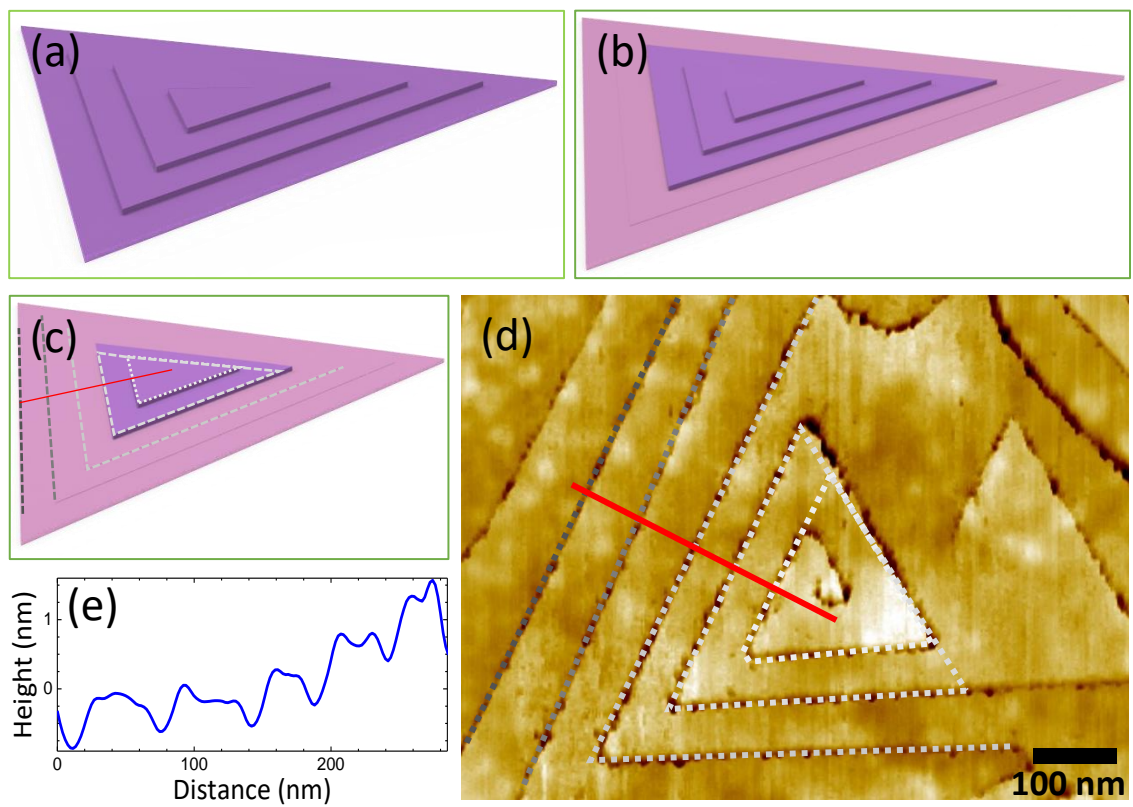

**Figure S4**: Schematic representation of the growth mechanism. (a) Only spirals and (b) both spirals and 2D nucleation with periodic defect lines present on a triangular domain. (c) A schematic showing a triangular domain marked to show spirals and 2D nucleation with IDPDs. (d) STM image showing both spirals and 2D nucleation with IDPDs. (e) A height profile drawn across line (in red) in the STM image confirms the schematic representation of the growth mechanism.